\documentclass[useAMS,usenatbib]{mn2e}
\pdfoutput=1
\title[Binary accretion rates: dependence on temperature and mass-ratio]{Binary accretion rates: dependence on temperature and mass-ratio}
\author[Young and Clarke]{
M.D.~Young$^1$, 
and
C.J. Clarke$^1$ \\
$^1$Institute of Astronomy, University of Cambridge, Madingley Road, Cambridge,
CB3 0HA, United Kingdom\\
}
\usepackage{graphicx}
\usepackage{natbib}
\usepackage{url}
\usepackage{amsmath}
\usepackage{amssymb}
\usepackage{caption}
\usepackage{subcaption}

\newcommand{\abs}[1]{\left| #1 \right|} 
 
\let\baraccent=\= 
\renewcommand{\=}[1]{\stackrel{#1}{=}} 

\newcommand{\mathv}[1]{\ensuremath{\mathbf{#1}}}

\begin{document}
\date{Written May 21st 2015}
\maketitle
\begin{abstract}

  We perform a series of 2D smoothed particle hydrodynamics (SPH) simulations
  of gas accretion onto binaries via a circumbinary disc, for a range of gas 
  temperatures and binary mass ratios ($q$).  We show that increasing the gas 
  temperature increases the accretion rate onto the primary for all values 
  of the binary mass ratio: for example, for $q=0.1$ and a fixed binary
  separation, an increase of normalised sound speed by a factor of $5$ (from
  our ``cold'' to ``hot'' simulations) changes the fraction of the accreted
  gas that flows on to the primary from $ 10\%$ to $\sim40\%$.  We present a simple 
  parametrisation for the average accretion rate of each binary component accurate 
  to within a few percent and argue that this parameterisation (rather than those in the
  literature based on warmer simulations) is relevant to supermassive black
  hole accretion and all but the widest stellar binaries.  We present
  trajectories for the growth of $q$ during circumbinary disc accretion and
  argue that the period distribution of stellar ``twin'' binaries is strong
  evidence for the importance of circumbinary accretion.  We also show that 
  our parametrisation of binary accretion increases the minimum mass ratio
  needed for spin alignment of supermassive black holes to $q \sim 0.4$, 
  with potentially important implications for the magnitude of velocity kicks
  acquired during black hole mergers.

\end{abstract}

\section{Introduction}

Understanding the accretion of material onto binaries is important for many
astrophysical problems.  Super massive black hole binaries are thought to
occur frequently as a result of large galactic mergers.  However,
observationally detecting black hole binaries remains a challenge 
\citep{BlackHoleSignatures,Dotti12,BHReview,BHObs}.  As such, understanding
the electromagnetic signatures of gas flows onto binaries is important for improving 
detection .  The relative accretion rate of gas by the 
two binary components is also important in setting the alignment of black hole spins
\citep{BlackHoleAlign1,BlackHoleAlign2,Gerosa}.  The alignment of black
hole spins sets the recoil velocity of the post-merger black hole and can
influence the host galaxy through feedback processes \citep{BlackHoleGrowth}. 
Each black hole's spin is aligned with its disc by the Bardeen-Petterson
effect \citep{BP}, on a time scale which depends upon the accretion rate through 
the disc \citep{Gerosa}.  As such, understanding the 
relative strength of primary ($\dot{M_1}$) and secondary ($\dot{M_2}$) 
accretion is vital for determining the likelihood of spin alignment.

The distribution of stellar mass within binaries is also sensitive to the
details of binary accretion.  Given that a large fraction of stars are 
in binaries, this problem is of great significance
\citep{BfracObs1,BfracObs2,BfracObs3,BfracObs4,BfracObs5,BfracObs6}.  It is well 
known that the initial mass of a protobinary is only a fraction of the mass of
the parent cloud (except for the widest binaries)
\citep{BossSeedMass,BBSeedMass}.  As most of the mass not in
the initial protobinary is of greater specific angular momentum, the
protobinary can only continue to grow by accreting material through a
circumbinary disc.  The ultimate mass ratio of the binary ($q=M_2/M_1$, where
subscripts $1$ and $2$ refer to the primary and secondary, respectively) is
set by the accretion rates onto the primary ($\dot{M_1}$) and secondary
($\dot{M_2}$) through the circumbinary disc.  

It is possible that the circumbinary disc that feeds material onto the
binaries may be misaligned with the binary orbit.  Accretion via a misaligned
circumbinary disc does not change the component of the binary that accretes
the most material \citep{Nixon11,Nixon13,Dunhill14}, although the accretion
rate is potentially higher for a misaligned circumbinary disc \citep{Nixon13}.

Cluster level simulations have been used to predict properties of
binaries \citep{Goodwin_ClusterSim1,Offner_ClusterSim,Bate_ClusterSim,Delgado_ClusterSim}.
They find mass distributions peaked towards high values of
$q$, implying preferential accretion onto the secondary.  Dedicated studies of
isolated binaries, which allow for much greater resolution, have also been 
performed using both smoothed particle hydrodynamics (SPH) in 3D
\citep{BateSPH,Dotti10} and 2D \citep{MYbinary},
and using grid codes in 2D \citep{Hanawa,TTauriGrid,Farris14}.  The overwhelming majority
of these studies find a strong tendency for the majority of the mass to flow
onto the secondary.  This is because material flowing from the circumbinary
disc encounters the secondary first, since its orbit is further from the
centre of mass of the system.  Indeed, the same preference for accretion onto
the secondary can be found in simulations of ballistic (non-collisional)
material with specific angular momentum greater than the binary \citep{BateBallistic}.

However, simulations have also shown that gas temperature can 
moderate this preference for accretion onto the secondary
\citep{Dotti10,MYbinary}.  This is because
at higher temperatures, material travels from the circumbinary disc onto the
binary along a wider range of trajectories, allowing some of it to orbit the 
edge of the secondary's Roche lobe and accrete onto the primary
\citep{MYbinary}.  Understanding 
the dependence of relative accretion rate on both gas temperature and binary
mass ratio is vital for providing accurate predictions of black hole spin
alignment and the distribution of $q$ in stellar clusters.  Understanding the
temperature dependence of accretion rates is particularly important for black
hole applications, where the ratio of sound speed to binary orbital speed is
too low to be modeled by current simulations \citep{Gerosa}.

In this paper we present a series of SPH calculations of the accretion of gas
onto binaries at a range of gas temperatures and mass ratio.  We use our 
simulations to propose a simple parametrisation of the
relative accretion rate of binaries, as a function of both gas temperature and
mass ratio. In Section \ref{sec:model} we describe the details of our model 
and the computational set up.  In Section \ref{sec:results} we present the 
results of our simulations.  In Section \ref{sec:discuss} we discuss the 
interpretation of our results, their implications. Finally we provide 
some concluding remarks in Section \ref{sec:conclude}.

\section{Model details}
\label{sec:model}

We model a binary with mass ratio $q=M_2/M_1$, separation $a$, total mass
$M=M_1+M_2$ on a anti-clockwise circular orbit.  All simulations are performed 
in 2D using a modified version of the SPH code GADGET2 \citep{Gadget2Code}.  We 
choose to use 2D rather than 3D for comparison to earlier work \citep{MYbinary} and to
reduce computational costs (for $N$ particles, resolution scales as $N^{-1/2}$
in 2D and $N^{-1/3}$ in 3D).

Physically, accretion onto a binary is mediated by a circumbinary accretion
disc.  As viscosity in this disc causes it to spread inwards, material flows
from the edge of the circumbinary disc onto the binary.  Self consistent
modelling of this viscous spreading, which is primarily driven by an effective
viscosity resulting from turbulent processes \citep{PringleDiscs,SS73}, 
is computationally
challenging for binary accretion modeling and has not to date been extended to
studying accretion onto the individual binary components \citep{fullmri}.  
Furthermore, the details of the accretion 
in the circumbinary disc are unlikely to affect the steady-state accretion rate 
of the binary.  Given this, we choose to follow \cite{BateSPH} \&
\cite{MYbinary} and inject 
material at a fixed boundary far from the binary, $R=R_{inf}a$, with fixed specific 
angular momentum, $j=j_{inf}\sqrt{GMa}$.  The initial radial velocity of material at
the outer edge is set such that the material is initially marginally
gravitationally bound.  This allows us to provide a continuous
supply of material to the binary, without the computational expense of
simulating a large circumbinary disc.

The mass of the gas is taken to be negligible compared to the binary and so
the self-gravity of the gas is ignored.  For the same reason, the binary 
masses and orbits remain fixed throughout our simulations. Any particle 
which is within $R=R_{acc}a$ of a binary particle is removed from the
simulation and the number of removed particles per binary particle is
recorded (we set $R_{acc} = .02$ in all our simulation).  The gas temperature 
is modeled using an isothermal equation of
state, which is parametrised in terms of c:
\begin{equation}
  c = c_s / \sqrt{\frac{GM}{a}}
  \label{eq:cparam}
\end{equation}
$c$ is then the sound speed normalised to the orbital speed of the binary
(it is also $H/R$ of the circumbinary disc at a radius of $a$).  

For all simulations performed here we set $j_{inf}=1.2$ and $R_{inf}=20$ which
ensures a constant supply of material onto the binary with specific angular 
momentum similar to what would be expected for material at the inner edge of a
circumbinary disc \citep{BateSPH,Bate_props}.  Lower values of $j_{inf}$ would 
be unrealistic for accretion via a disc, while higher values of $j_{inf}$ 
would cause a disc to form at a radii where material cannot flow onto the 
binary until viscous
spreading moved the boundary inwards.  We inject particles at the boundary at
a rate of $\dot{N}=500$ per dynamical time, which ensures the edge of the
circumstellar discs are well resolved \citep{MYbinary}.  As the temperature of 
the gas  has been shown to affect the relative accretion rate of binary components, 
we explore values in the range $c=0.05 - 0.25$.  As lower values of $c$
produce thinner discs which require more resolution elements to resolve, we
were unable to probe values of $c < 0.05$.  We run simulations with
$0.1<q<1.0$ to explore the effect of $q$ on relative accretion rates.  Smaller
values of $q<0.1$ were deliberately excluded as they are more computationally
demanding and the nature of the accretion flow has been shown to change
qualitatively in this range.  That is, for $q<0.1$, the accretion transitions
from gas flowing onto a binary in a low gas density cavity, to gas flowing
through a large circumprimary disc, with a large embedded body clearing a gap
\citep{GapFlow}.  Each simulation was run for at least $500$ binary dynamical times,
by which time they have settled into a steady-state \citep{MYbinary}.

\begin{figure}
  \begin{center}
	 \includegraphics[width=0.5\textwidth]{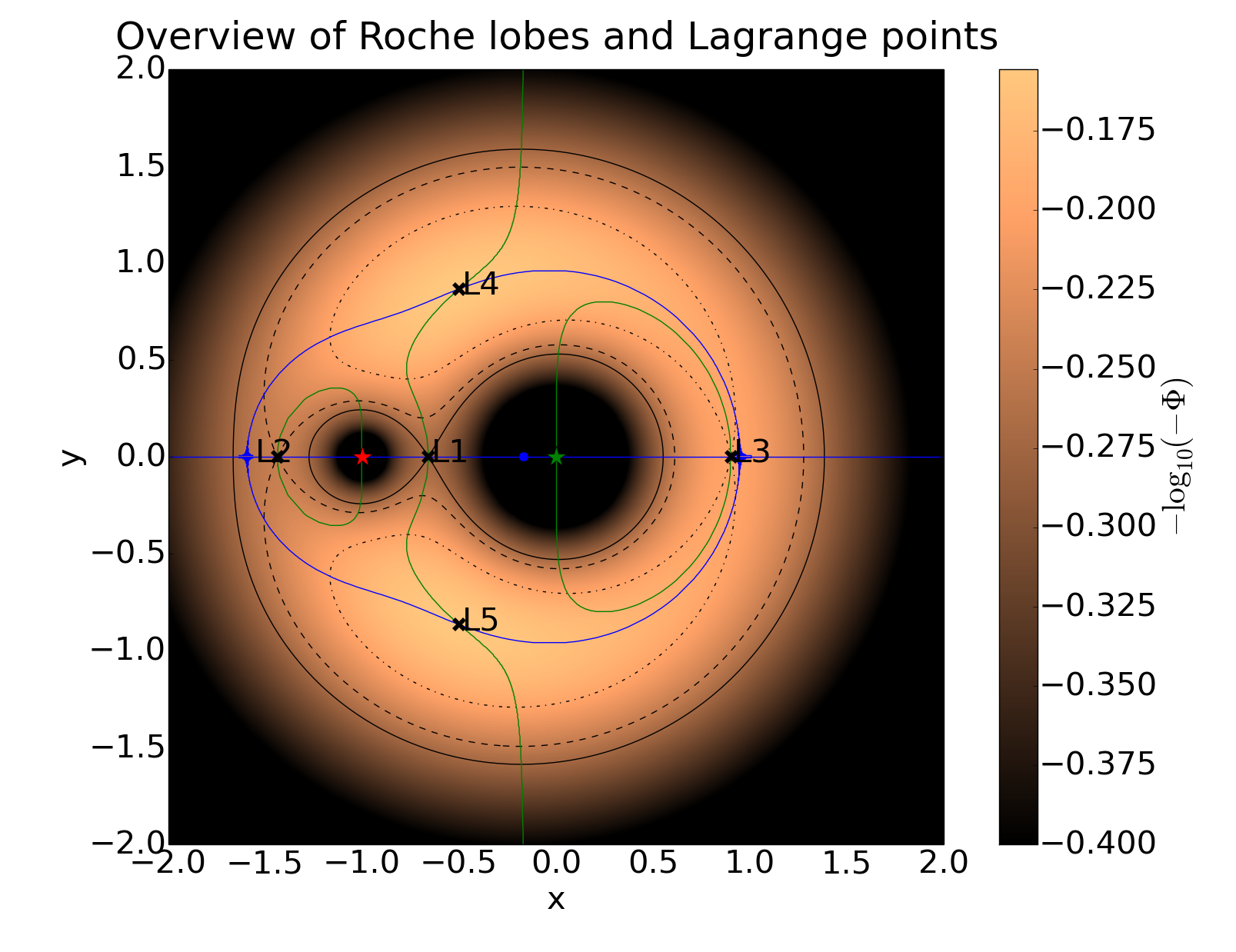}
  \end{center}
  \caption{Equipotential surfaces of $\Phi$, the modified potential
  (gravitational potential plus term to include centrifugal force 
  of non-inertial frame) for the values of $\Phi$ obtained at the three
  Lagrange points L1,L2 and L3.  The colour map shows $\Phi$ elsewhere.  The
  stars are marked in red (secondary) and green (primary) and the centre of
  mass of the system is marked in blue.  The green and blue lines are the
  locations where the x and y components of the force resulting from this
  potential (i.e., $-\nabla{\Phi}$) are zero.}
  \label{fig:Roche_lobes}
\end{figure}

All analysis was performed in a frame corotating
with the binary, translated so that the primary is at the origin and the
secondary is at (x,y)=(-1,0).  In this reference frame it is possible to
define a modified potential, commonly called the ``Roche potential'', as:
\begin{equation}
  \Phi = -\frac{GM_1}{\abs{\mathv{r}-\mathv{r_1}}}
  -\frac{GM_2}{\abs{\mathv{r}-\mathv{r_2}}}
  -\frac{1}{2}(\mathv{\Omega} \times \mathv{r})^2
  \label{eq:roche}
\end{equation}
where $\mathv{\Omega}$ is the angular momentum vector of the binary, $\mathv{r}$ is
the position vector and subscripts $1$ and $2$ refer to the primary and
secondary respectively.  Surfaces of equipotential of $\Phi$ and Lagrange
points (points of zero gradient) are shown in Figure \ref{fig:Roche_lobes}.

Table \ref{tab:simulations} lists the parameters used for all simulations in this paper.
Details of the exact version of the SPH code used along with initial
conditions and parameter files can be found in Section \ref{sec:methods}.

\begin{table}
  \centering
  \begin{tabular}{|l|l|l|l|}
    ID & $q$ & $c$ & $t_{end}$ \\
    \hline
    1 & 0.1 & 0.05 & 720 \\
    2 & 0.2 & 0.05 & 500 \\
    3 & 0.3 & 0.05 & 640 \\
    4 & 0.4 & 0.05 & 630 \\
    5 & 0.5 & 0.05 & 690 \\
    6 & 0.6 & 0.05 & 720 \\
    7 & 0.7 & 0.05 & 730 \\
    8 & 0.8 & 0.05 & 840 \\
    9 & 0.9 & 0.05 & 800 \\
    10 & 0.1 & 0.25 & 2000 \\
    11 & 0.2 & 0.25 & 2000 \\
    12 & 0.3 & 0.25 & 2000 \\
    13 & 0.4 & 0.25 & 2000 \\
    14 & 0.5 & 0.25 & 2000 \\
    15 & 0.6 & 0.25 & 2000 \\
    16 & 0.7 & 0.25 & 2000 \\
    17 & 0.8 & 0.25 & 2000 \\
    18 & 0.9 & 0.25 & 2000 \\
    19 & 0.1 & 0.10 & 542 \\
    20 & 0.1 & 0.15 & 521 \\
    21 & 0.1 & 0.20 & 515 \\
  \end{tabular}
  \caption{Parameters used for simulations in this paper.  For each
  simulation particles were injected at a rate of roughly $\dot{N}\approx 500$
  per dynamical time ($\sqrt{a^3/GM}$), the accretion radius was set to
  $0.02a$, particles were injected from $20a$ with specific angular momentum
  $1.2\sqrt{GMa}$.}
  \label{tab:simulations}
\end{table}

\section{Results}
\label{sec:results}

All our simulations undergo a transient settling phase, which is dependent
on the details of the initial conditions, before settling into a steady-state
after $\sim 30$ binary orbits.  To exclude the physically uninteresting 
effects of this settling phase, we excluded the first 200 dynamical times
(where $2\pi$ dynamical times is one binary orbit) from all the results that follow.  

The evolution of the binary mass ratio $q$ and the relative
strength of primary/secondary accretion flows can be characterised using the 
parameter,
\begin{equation}
  \Gamma = \frac{\dot{q}}{q}/\frac{\dot{M}}{M} =
  \frac{(1+q)}{q\dot{M}}(\dot{M_2}-q\dot{M_1})
  \label{eq:Gamma}
\end{equation}
which is the fractional change in $q$ per fractional change in $M$.  
This can be re-written as,
\begin{equation}
  \Gamma = \frac{(1+q)}{q} (1-\lambda (1+q))
  \label{eq:GammaSPH}
\end{equation}
where $\lambda = \dot{M_1}/\dot{M}$ and $\dot{M} = \dot{M_1}+\dot{M_2}$.
In our simulations we measure $\dot{M_1}$ and $\dot{M_2}$ by counting 
the number of particles accreted onto the primary and the secondary.  

\begin{figure}
  \begin{center}
	 \includegraphics[width=0.5\textwidth]{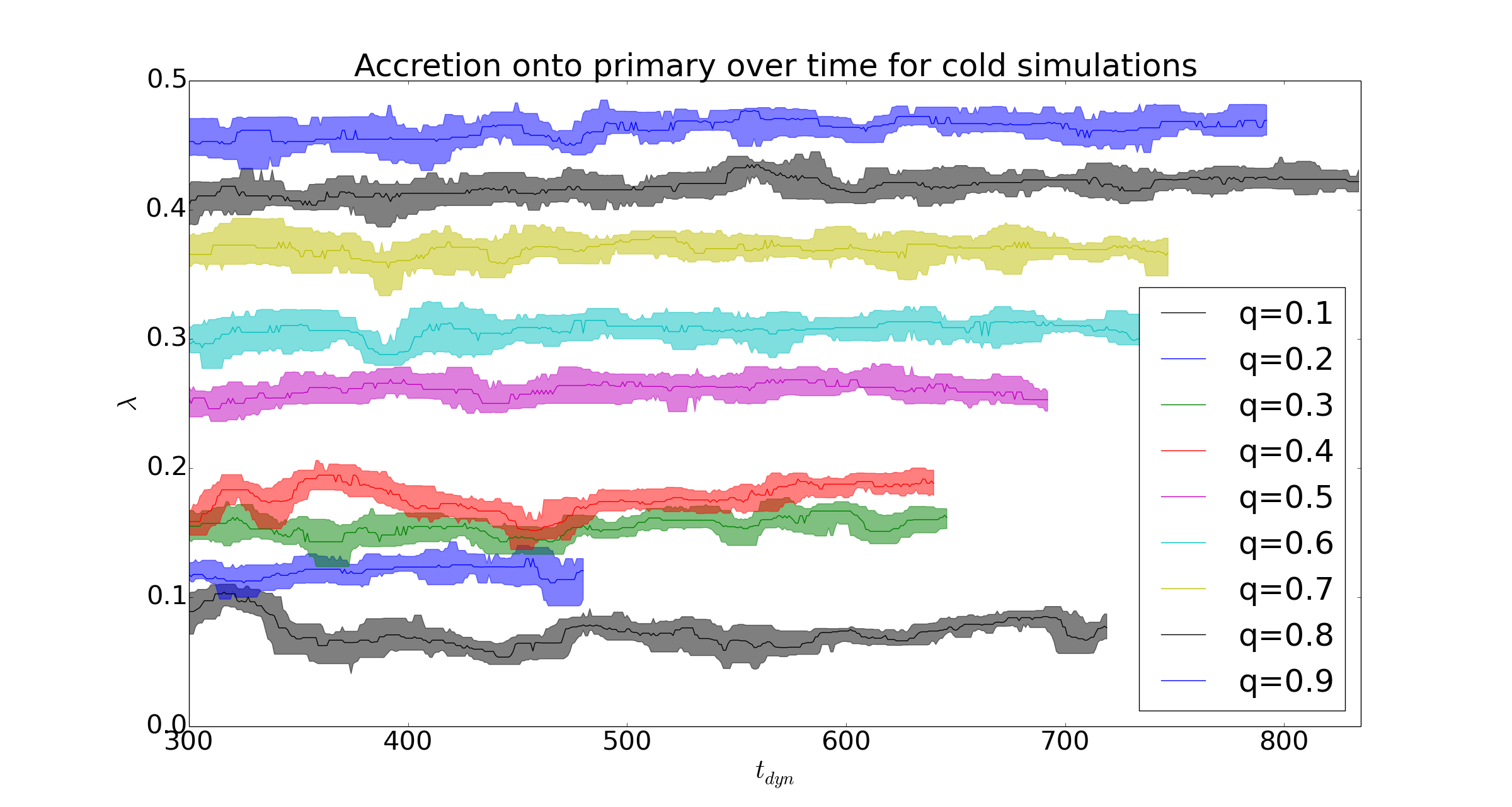}
  \end{center}
  \caption{Steady-state values of $\lambda=\dot{M_1}/\dot{M}$ for simulations
  with cold gas ($c=0.05$), for a range of values of $q$.  The central line
  shows a moving average over 3 binary orbits of $\lambda$ while the shaded
  region shows the one standard deviation range on $\lambda$ within this
  range.  The first 200 binary dynamical times are omitted to ensure that only
  steady-state accretion is shown.}
  \label{fig:coldlam}
\end{figure}

The evolution of $\lambda$ as a function of time and $q$ is shown for
simulations with cold gas ($c=0.05$) in Figure \ref{fig:coldlam}.  $\lambda$ is smoothed
using a moving average over three binary orbits, which is roughly the orbital
period of the edge of the binary cavity.

\begin{figure}
  \begin{center}
	 \includegraphics[width=0.5\textwidth]{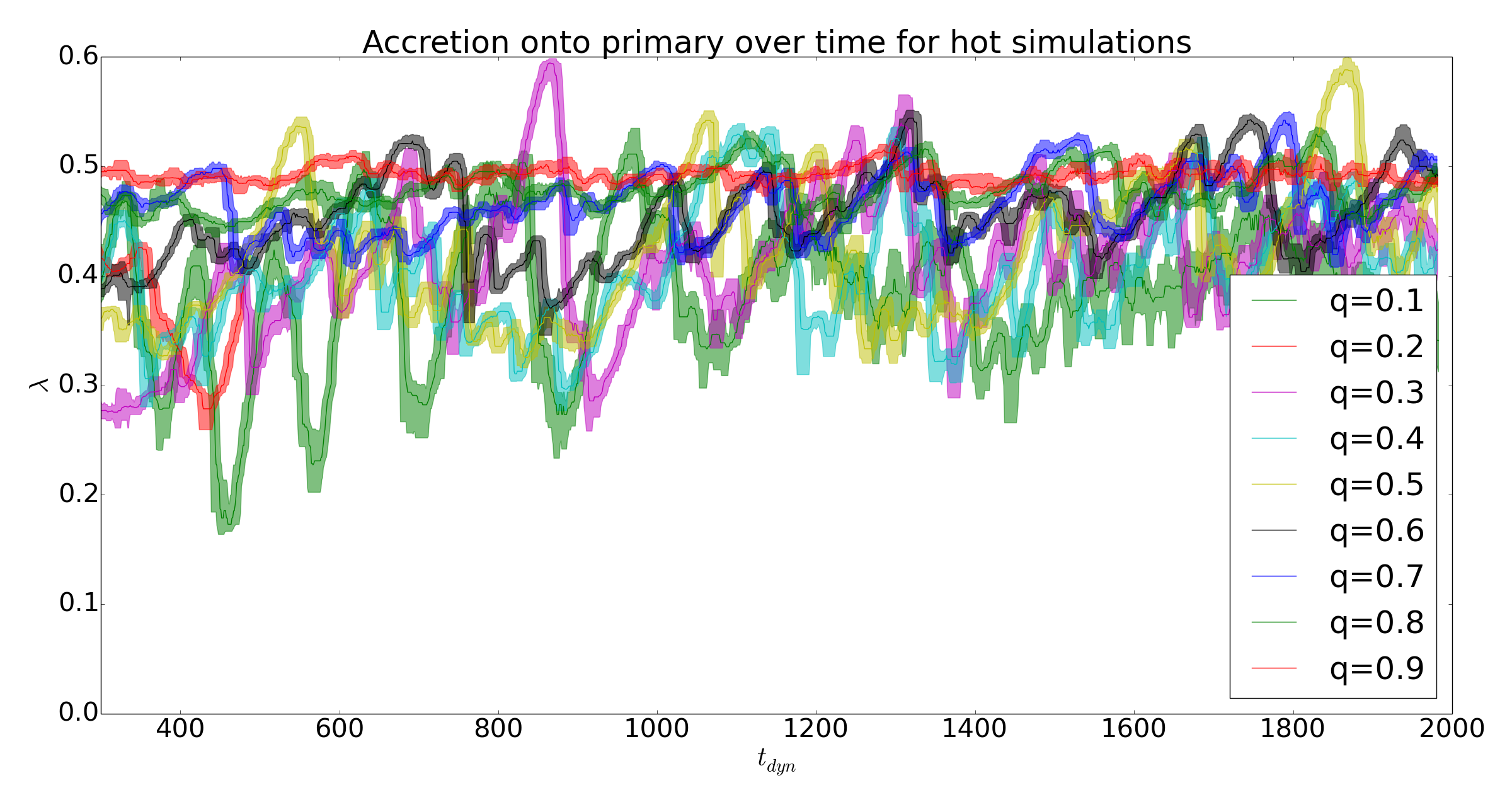}
  \end{center}
  \caption{The same as Figure \ref{fig:coldlam} but for simulations with hot
  gas ($c=0.25$).}
  \label{fig:hotlam}
\end{figure}

Figure \ref{fig:hotlam} shows the same information for the hot simulations (where
$c=0.25$).  Unlike the cold simulations the accretion rate $\lambda$ is
significantly more variable, making it difficult to discern any trend in the
average accretion rate with $q$.

\begin{figure}
  \begin{center}
	 \includegraphics[width=0.5\textwidth]{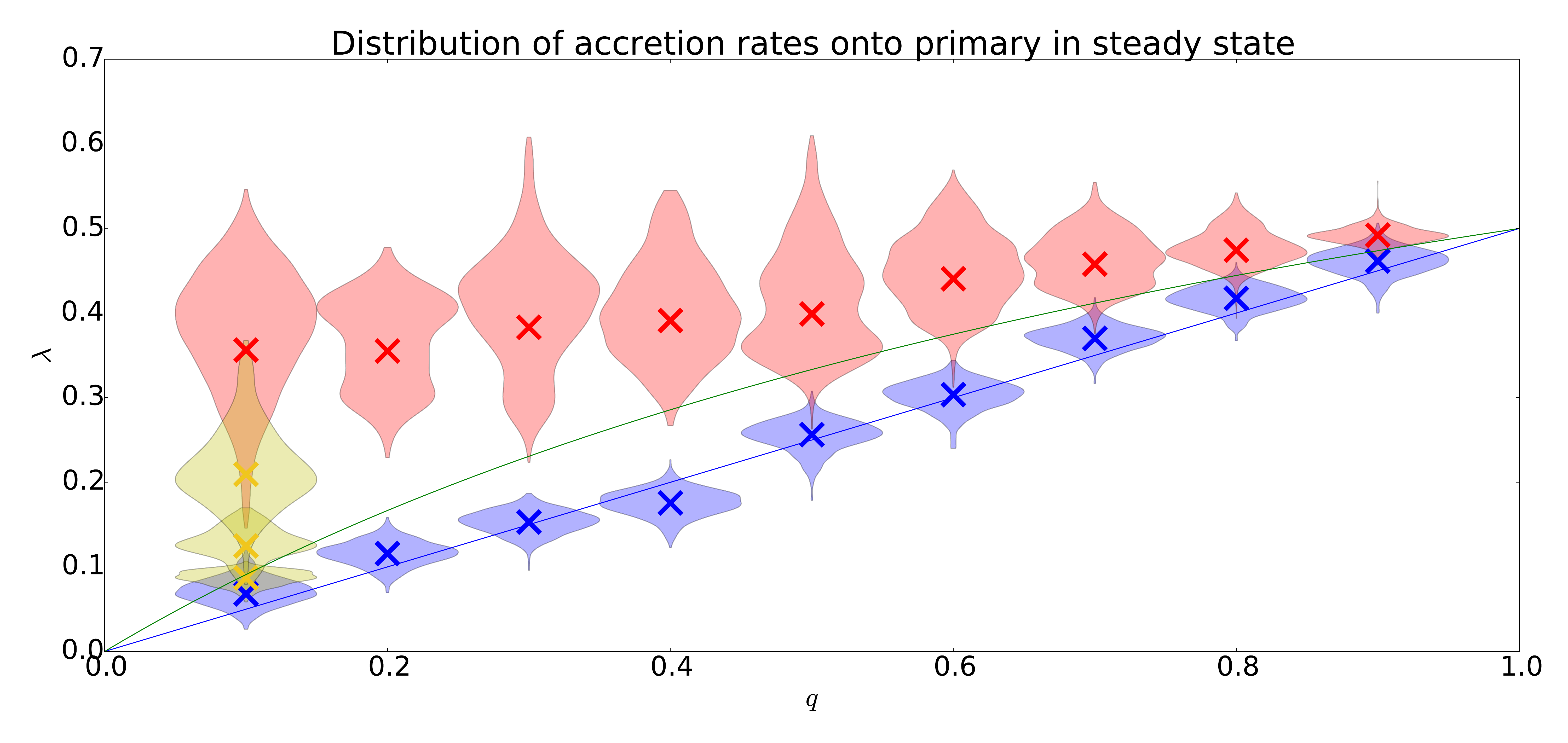}
  \end{center}
  \caption{The distribution of steady state $\lambda$ for all simulations.
  Each simulation is represented by a violin plot, where more common values of
  $\lambda$ correspond to a greater horizontal extent.  The crosses show the average
  steady-state accretion rate across the entire simulation (excluding
  $t<200t_{dyn}$), $\int \dot{M_1}/\int \dot{M}$.  Hot simulations are shown
  in red, cold simulations in blue and intermediate simulations in yellow.
  The blue line shows the parametrisation $\lambda=q/2$ and the green line
  shows $\lambda=\frac{q}{q+1}$}
  \label{fig:violins}
\end{figure}

Figure \ref{fig:violins} show the distribution of $\lambda$ in the steady-state for all
simulations.  This allows the trends in the average accretion rate 
with $q$ and $c$ to be more clearly visualised.  The cross in each
distribution shows an estimate of the average steady-state
$\lambda$, given by $\int \dot{M_1}/\int \dot{M}$.

Figure \ref{fig:violins} also shows the distribution of steady-state accretion rates for a
series of simulations with $q=0.1$ and gas temperatures intermediate between
hot ($c=.25$) and cold ($c=.05$).  These additional simulations show how the
accretion rate changes with gas temperature in more detail.

\section{Discussion}
\label{sec:discuss}

\subsection{Variability of $\lambda$}

All simulations show a significant amount of temporal variability in
$\lambda$.  This variability is driven by fluctuations in the amount of
material flowing into the binary cavity, which has been identified in many
previous simulation of gas flow onto binaries 
\citep{GapFlow,Ochi,MYbinary,BlackHoleSignatures,FlowIntoGap,fullmri}.

The physical mechanism for the variability in supply of material onto the binary was
explained in detail by \cite{FlowIntoGap} \& \cite{GapFlow}.  They found that the edge of the
circumbinary disc was eccentric (i.e., the cavity wall is not circular).
Because of this, the amount of material flowing inwards fluctuates on the time scale 
of, 1) the binary orbit and 2) the precession time scale of the disc edge.

\begin{figure}
  \begin{center}
    \includegraphics[width=0.5\textwidth]{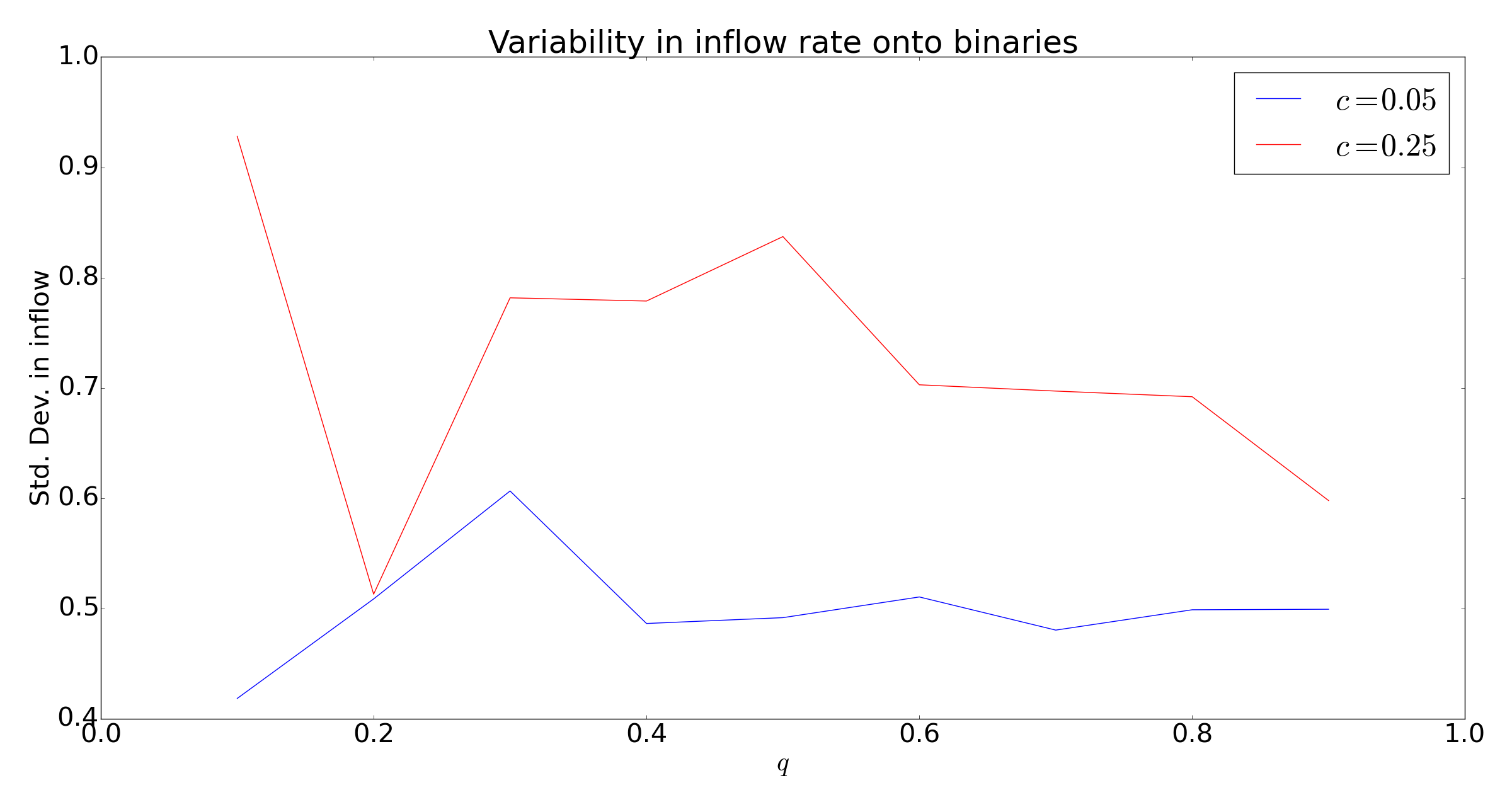}
  \end{center}
  \caption{The vertical axis is the standard deviation in the rate of inflow onto 
  either Roche lobe, divided by the particle injection rate at the outer
  boundary (i.e., the standard deviation of the inflow rate into the cavity).  
  As with all other quantities, the first 200 dynamical times were
  excluded from the calculation so that only the steady state is probed.}
  \label{fig:inflow}
\end{figure}

Because changes in the rate of flow onto the binary are driven by disc
eccentricity, the magnitude of the variability (and the importance of the
different time scales) varies with $q$.  Generally, a larger $q$ is able to
excite a more eccentric cavity edge, which leads to greater variability (see
\cite{GapFlow} for details of the physical mechanism).  Figure
\ref{fig:inflow} shows the variability in the amount of gas flowing inwards past 
the Roche equipotential surface passing through the L1 point. Although there is 
significant variability 
in the infall rate, the amount of variability does not change systematically with 
$q$.  This is likely a limitation of our decision to inject material with constant
angular momentum $j=1.2\sqrt{GMa}$, which prevents the disc edge reaching
high eccentricities (\cite{FlowIntoGap,fullmri} found $e \sim 0.1 - 0.4$ for
$q=1.0$) by providing a flow of material that will push the disc towards a
circular flow beyond the circularisation radius of $R=j_{inf}^2a$.  On the other 
hand, a trend towards increased variability at higher gas temperatures can clearly 
be seen.

Unlike Figure \ref{fig:inflow}, Figure \ref{fig:violins} shows a clear trend 
towards the accretion rate onto either component of the binary becoming \emph{less}
variable as $q$ increases.  This is because $\lambda$ (and the accretion onto
the secondary, $1-\lambda$) measures the accretion rate through the inner edge
of the discs around each component of the binary.  Material flows from the edge 
of the circumbinary disc onto either the circumprimary or circumsecondary disc.
To reach the inner edge and be accreted, it must be ``processed'' by the
circumstellar disc.  That is, material that joins the circumprimary disc is
only registered at the inner edge once that material has been moved there by
the circumprimary disc's viscosity.  For discs with lower viscosity (e.g.
colder discs) or with a greater spatial extent (e.g. discs with higher $q$),
variability in the supply of material at the outer edge will be more smeared
out by the time the inner edge is reached.

Given this, the variability in Figure \ref{fig:violins} is likely driven
largely by numerics, with simulations with hotter gas and/or larger discs
(i.e. higher $q$) transmitting more of the variability in supply of material
striking the discs around each binary component through to the inner edge where 
it is accreted.  Physically realistic discs, with significantly lower viscosity 
than the numerical viscosity of our simulations, greater dynamic range in disc
radius and lower gas temperatures, will have much less variability in the
accretion rate onto the components of the binary than our simulations.
However, the average accretion rate over many binary orbits should be
independent of the amount of numerical variability.  This can be seen in the
simulations of \cite{MYbinary}, where the magnitude of the variability decreases with
resolution but the average accretion rate is unchanged.

\subsection{Parametrisation of $\lambda$}

The most important quantity to understand for our purposes is the steady-state
value of $\lambda$, which uniquely determines the fractional accretion rate
$\Gamma$ via Equation \ref{eq:GammaSPH}.  For the reasons discussed in the
previous section, it is the steady state average of $\lambda$ which is
important in determining the physical properties of the binary.  

It was recently suggested by \cite{Gerosa}, that $\lambda$ can be parametrised 
as,
\begin{equation}
  \lambda = \frac{q}{q+1}
  \label{eq:lodatolambda}
\end{equation}
based upon the data of \cite{Farris14}, who simulated accretion
onto binaries using a grid-based code for a range of values of $q$ and
intermediate temperature gas ($c=0.10$).

The green line in Figure \ref{fig:violins} shows this parametrisation.  It is clear 
that this expression is appropriate for binaries accreting gas of intermediate 
temperature (understandable, given the source data), but provides a poor estimate 
for the accretion of hot or cold gas.

By symmetry, any parametrisation must have $\lambda=0.5$ when $q=1$.  Based 
on the data in Figure \ref{fig:violins}, we propose the simple parametrisation 
\begin{equation}
  \lambda = \frac{1}{2}+m(q-1)
  \label{eq:mylambda}
\end{equation}
where $m$ is a function of $c$ and $m \rightarrow 0.5$ as $c \rightarrow 0$.  The cold
limit, where $\lambda = q/2$, is shown in blue in Figure \ref{fig:violins} and is 
accurate for the cold simulations to within a few percent.  

\begin{figure}
  \begin{center}
   \includegraphics[width=0.5\textwidth]{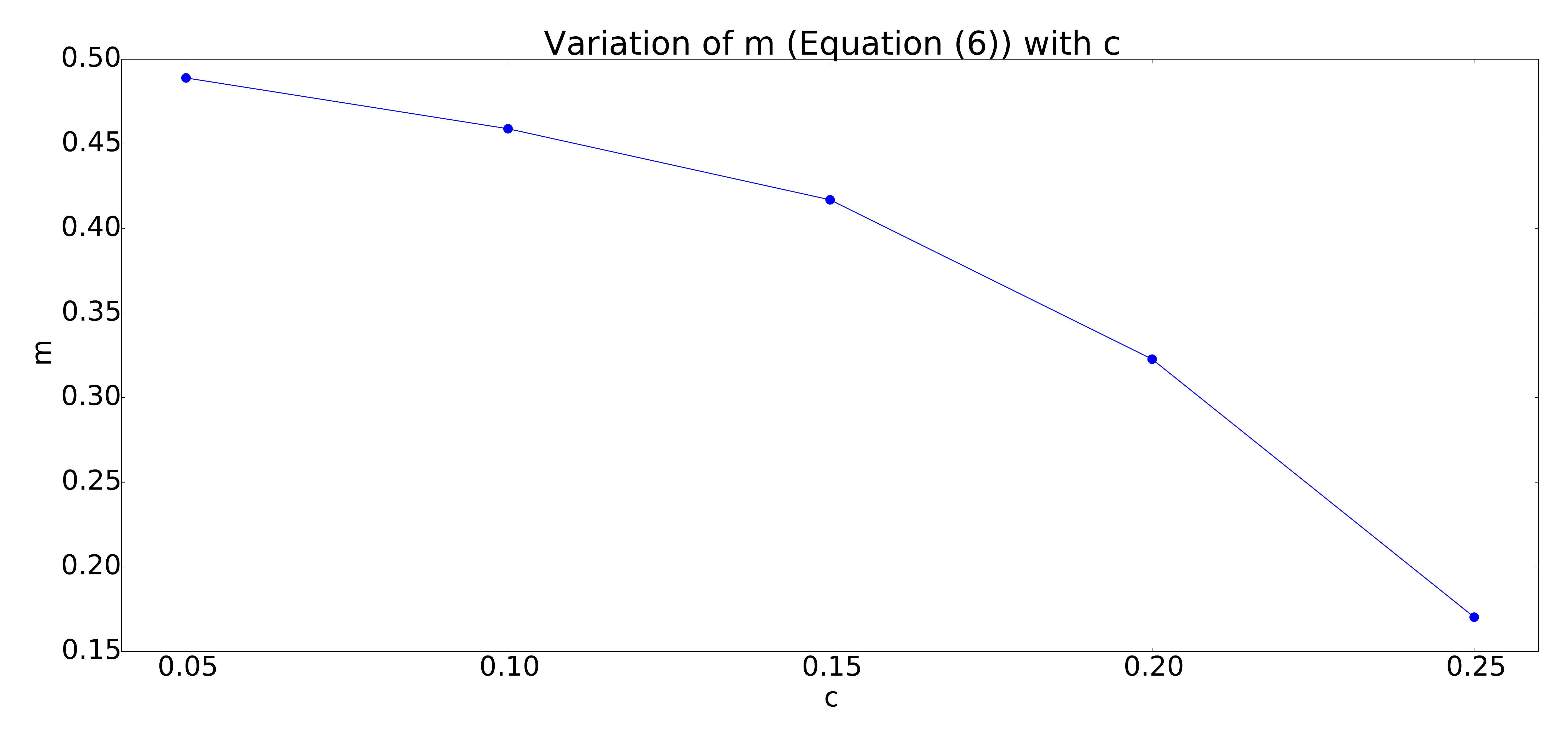}
  \end{center}
  \caption{The best fit value of the parameter $m$ in the parametric model for
  $\lambda$ given by Equation \ref{eq:mylambda}.  For each value of $c$
  Equation \ref{eq:mylambda} was fit using the average steady state value of
  $\lambda$ for each $q$.  Note that the values for $c=0.10$,$0.15$ \& $0.20$
  are fit using just one data point.}
  \label{fig:mgrow}
\end{figure}

To asses how $m$ in Equation \ref{eq:mylambda} varies with $c$, we calculated 
the best fit value of $m$ to the data using least squares minimisation, which we 
plot in Figure \ref{fig:mgrow}.  It is
clear that an asymptotic limit of $m=0.5$ is appropriate for the cold limit
and that $m$ decreases as a strong power of $c$.  We have avoided giving a
functional form for $m$ given that we have relatively few data points in $c$ 
and the values of $m$ for $c=0.10$,$0.15$ and $0.20$ are fit using only one 
sampling in $q$.

Moreover, it is the cold limit which is most applicable to the bulk of
astrophysical systems.  In the case of
protobinaries, $c$ is likely to lie in the range $\sim 0.01-0.1$
\citep{MYbinary}.  For black hole binaries, normalised sound speeds as low 
as $.001$ are possible \citep{Gerosa}.  In either case, our results 
suggest that the binary accretion rate should be well approximated by,
\begin{equation}
  \Gamma = \frac{(1-q^2)(2+q)}{2q}
  \label{eq:Gfinal}
\end{equation}
As well as being in good agreement with the results of Section
\ref{sec:results}, this parametrisation is also consistent with the finding 
of earlier simulations of cold gas accreting onto binaries \citep{BateSPH,MYbinary}.

\subsection{Implications}

For all values of parameter space that we explored, we find $\Gamma>0$, implying that
accreting material onto a binary always brings its mass ratio closer to $1$.  
It has been shown that when $q<0.1$, a region of parameter space we chose not
to explore, the physical behaviour of the accretion
flow transitions to that of a planet embedded in a disc, causing $\Gamma$ to
fall \citep{GapFlow,Farris14}.  It is possible that for very small values 
of $q$, $\Gamma$ may become negative, but this would need to be confirmed 
with specialised simulations.

\begin{figure}
  \begin{center}
	 \includegraphics[width=0.5\textwidth]{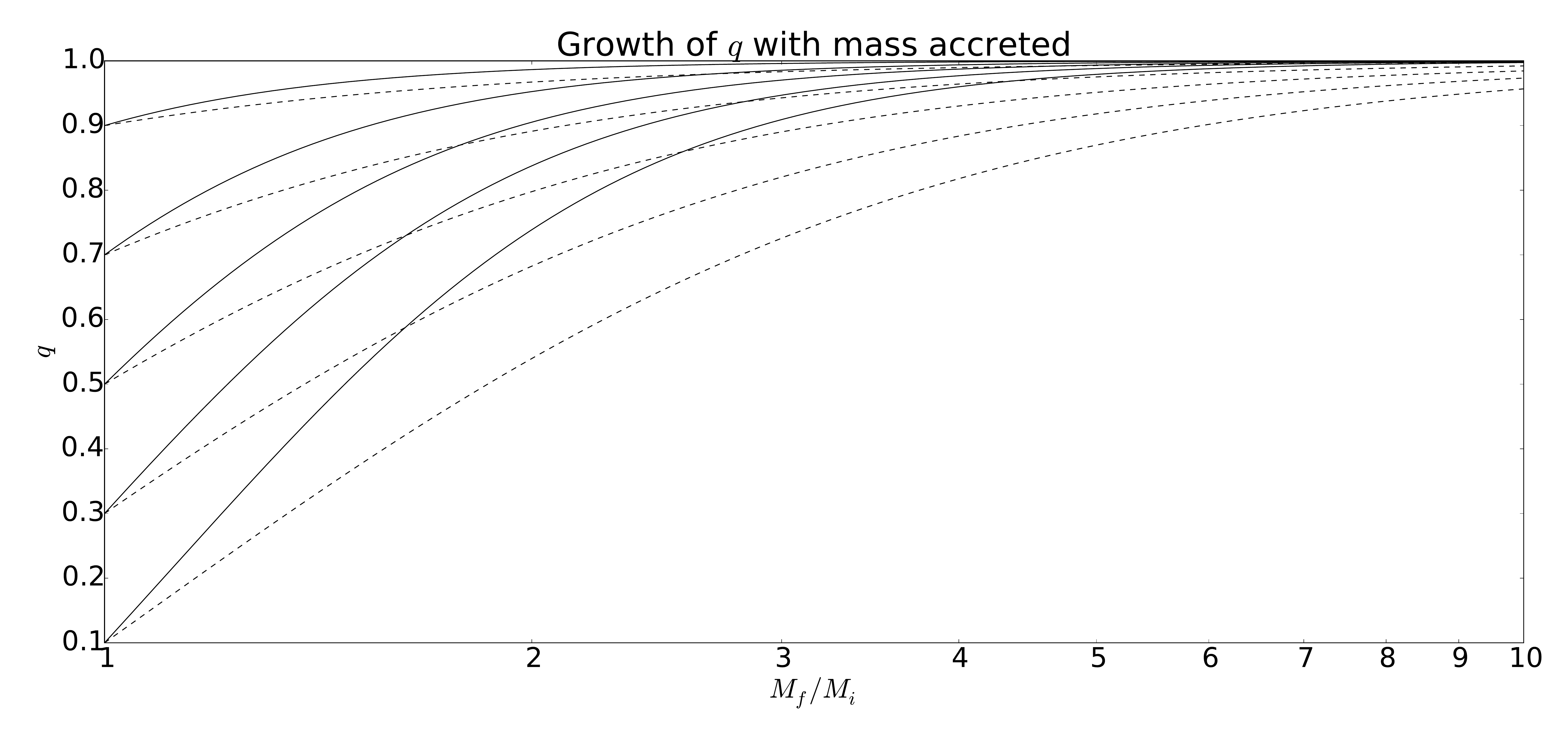}
  \end{center}
  \caption{Change in $q$ as a function of accreted mass over initial binary
  mass, for cold gas with $\lambda=q/2$ (solid line) and hot gas with
  $\lambda=0.17q+0.33$ (dashed line).}
  \label{fig:qgrow}
\end{figure}

By integrating Equation \ref{eq:Gfinal}, we can determine how $q$ evolves as a
function of accreted mass.  This is shown for both hot and cold simulations (dashed 
and solid lines, respectively) in Figure \ref{fig:qgrow}.  Note that this plot
is somewhat different from those presented in \cite{Bate_props}, in which a
protobinary was evolved subject to infall from a variety of parent cores.  The
resulting evolution in $q$ was then sensitive not only to the density and
angular velocity profile of the cores, but also to the assumptions made about
protobinary migration.  Specifically, $\dot{M_1}/\dot{M_2}$ depends on the
specific angular momentum of the infalling gas {\it relative to the binary},
which changes as the binary migrates.  Here we define the initial binary as
corresponding to the point at which it begins to accrete from a circumbinary
disc.  From this point on, the material accreted onto the binary must always
be of similar specific angular momentum to the binary and so $q$ simply
evolves along the appropriate path in Figure \ref{fig:qgrow}.  Figure
\ref{fig:qgrow} is therefore readily applicable to either the stellar binary
or black hole case.

\subsubsection{Application to stellar binaries}

Main sequence binary star surveys of solar type stars reveal a remarkable
population of ``twin-like'' binaries ($q > 0.95$; \cite{q_dist_disn})
\footnote{The existence of these ``twin-like'' binaries is not in doubt,
however, there may be some under-representation of the lowest $q$ binaries
as their abundance must be corrected in surveys which are incomplete
\citep{GoldbergComplete}.}. In the
light of Figure \ref{fig:qgrow} and given the lack of obvious mechanisms that
otherwise produce such strongly correlated masses, it is natural to ascribe
these objects to mass equalisation by circumbinary disc accretion.  Figure
\ref{fig:qgrow} implies that such twin-like systems can be produced from a
wide range of initial $q$ values, provided that the binary accretes of order
several times its own mass from a circumbinary disc.  On the other hand 
studies of binary migration during circumbinary disc accretion all concur 
that the (inward) migration time is less than or of the same order as the
mass doubling time \citep{FlowIntoGap,fullmri,BlackHoleMergers} . This 
immediately implies that ``twin-like'' systems have migrated inwards by a 
large factor. A further corollary is that
the parent system (binary plus circumbinary disc) would in the past
have contained an angular momentum that is considerably larger
than that contained in the present day binary\footnote{The simulations
presented here have been artificially set up so that the specific angular
momentum of the infalling material only modestly exceeds that of
the binary in order to mimic the inner parts of a circumbinary disc
in a computationally efficient manner; in reality, however,
the inner regions of such a disc are fed by viscous redistribution
within a reservoir of material whose mean specific angular momentum 
far exceeds that of the binary and this becomes increasingly true 
as the binary migrates inwards.}.

The period distribution of ``twin-like'' binaries amply supports this picture.
The angular momentum budget of pre-stellar cores \citep{MaxCloudJ}
is only sufficient to form binaries out to separations of order $10^3$ A.U.
\footnote{Wider binaries are attributed either to cluster mediated capture
\citep{WideBinaryFormation1,WideBinaryFormation2} or to
orbital reconfiguration in multiples producing wide companions on
high eccentricity, low angular momentum orbits \citep{WideBinaryFormation3}.} 
Given that the angular momentum in the final binary is only a fraction of the
angular momentum needed to drive a binary into a twin-like state, it is unsurprising
that the maximum separation of twin-like objects is considerably
less than this ($\sim 100$ A.U.).  On the other hand, low $q$
systems are relatively uncommon at small separations ($< 5$ A.U.); this
again is consistent with a picture in which the bulk of close
binaries are driven to small separation by circumbinary accretion
(rather than simply arising from cores with exceptionally low
specific angular momentum).

Figure \ref{fig:qgrow} also shows that more accreted mass is needed to
equalise the mass of binaries when the gas is ``hot''.  As the normalised
sound speed and seed protobinary mass as a fraction of total core mass both increase 
with binary separation  \citep{MYbinary,BossSeedMass,Bate_props}, binaries evolve 
less far to the right  (as less $M_f/M_i$ gas mass is available to be accreted) and 
along slower tracks (as the gas is ``hotter'') in Figure \ref{fig:qgrow}.  As such, 
we expect small separation binaries to have $q$ distributions strongly biased 
towards $q=1$, while wide binaries should show a much weaker preference for equal mass
binaries, again consistent with the picture described above.

\subsubsection{Black hole binaries}

\cite{Gerosa} examined whether circumbinary accretion onto
black holes can align their spins with the circumbinary disc on the
timescale on which they are driven inwards by circumbinary disc torques.
This analysis involved prescriptions for total binary accretion
rate, disc mediated migration, accretion induced spin alignment and
the relative accretion rates onto the binary components as a function
of $q$. The latter prescription was derived from the simulations
of \cite{Farris14}. In conjunction with the other model assumptions
it was concluded that whereas the secondary always aligns with
the circumbinary disc, the primary remains misaligned for $ q < 0.2$.
Misaligned spins at the point of binary merger have
important consequences for the recoil velocities resulting
from the merger event.

If we repeat this analysis but instead use our prescriptions
for differential mass acquisition (which  are  more appropriate for
the `cold' conditions expected in the black hole case) we find that
it is easier for black hole spins to remain misaligned: for the
fiducial model this increases the maximum $q$ value for misalignment
from $q=0.2$ to $q=0.4$. This is because we find a stronger
preference for accretion on to the secondary than in the warmer
\cite{Farris14} simulations; consequently the primary is relatively 
starved of accretion and can remain misaligned at higher $q$ values. 
Our result thus strengthens the result of \cite{Gerosa}, implying 
a significant population of black holes whose spins are expected to 
be misaligned at the point of merger.

When black holes with misaligned spins merge, the merger product 
receives a velocity ``kick'', the maximum value of which scales as
$\eta=q/(1+q)^2$ \citep{KickScaling}.  For binaries with $q \gtrsim 0.4$ the
maximum merger kick velocity may be comparable with the escape velocity of
typical elliptical galaxies \citep{RecoilEscapeVel}.  Moreover, the fact that
binaries may enter the gravitational wave inspiral regime with misaligned
spins and relatively large $q$ has potential implications for the richness of
the gravitational wave signature potentially detectable by e-LISA.

There is however a {\it caveat} to this. The prescriptions of
\cite{Gerosa} imply that the binary mass doubling timescale is much greater
than its migration timescale and in this limit it is appropriate
to consider spin alignment at constant $q$. However, recent
work employing `live' MRI turbulence in the gas flow finds a mass doubling
time which is  only modestly in excess of the migration time
\citep{fullmri}. If this is correct then migration by a large factor
(i.e. the $\sim 10$ fold shrinkage that is required before
migration due to gravitational wave emission becomes effective),
also implies a significant change in binary mass (and thus,
by Figure \ref{fig:qgrow}, the value of $q$). Indeed we
have seen above that in the stellar case there is ample
observational evidence for a preference for high $q$ pairs at small
separation. It is not clear in the black hole case whether
there is a sufficiently large mass reservoir in the
circumbinary disc to effect a significant change in $q$
and it is also unclear whether (for a restricted mass reservoir)
the relatively strong accretion on to the binary limits how
far it will migrate. Pending further theoretical work in this
area, we cannot rule out the possibility that even low q pairs
might not eventually attain a high enough value of q for
the primary to also align with the circumbinary disc.

\section{Conclusions}
\label{sec:conclude}

In this paper we have shown that the mass accretion rates of the primary and
secondary components of binaries depend on the gas temperature.  We have
shown that for all values of $q$, increasing the gas temperature allows more 
material to accrete onto the primary component of the binary. The relative 
accretion by the primary, $\dot{M_1}/\dot{M}$, is also found to increase at 
least quadratically with the sound speed of the accreted gas.

Although our simulations show a range of levels of variability of accretion
onto the star, we argue that it is only the higher than physical
disc viscosity and lower than physical dynamic range in disc radii that allows
variability in the mass influx at the Roche lobe to be manifest as variable
accretion onto the star. For realistic physical discs any variability in the inflow rate
will be smoothed out by the circumprimary/secondary disc and the variability
in the accretion rate onto the binary components will be low.  As such, it is
the average steady-state accretion rate that is the physically relevant
quantity in our simulations.
 
We have proposed a simple parametrisation of the relative accretion rates onto
the two components of the binary and shown that it reduces to,
\begin{equation}
  \frac{\dot{M_1}}{\dot{M}} = \frac{q}{2}
\end{equation}
and
\begin{equation}
  \frac{\dot{M_2}}{\dot{M}} = \frac{2-q}{2}
\end{equation}
when the accreted gas is cold.  

We argue that the period distribution of ``twin-like'' stellar binaries
provides strong evidence that these objects acquired the bulk of their
mass through circumbinary accretion and that they have migrated inwards
by a large factor from their initial birthplaces. Applying our parametrisation 
to super massive black hole binaries
with existing models for spin alignment we find that black hole spins' 
remain unaligned when $q$ is less than $0.4$.  However, this value may be
decreased if the mass accreted during the black hole merger is sufficient to
drive significant evolution in $q$.

\section{Materials \& Methods}
\label{sec:methods}

In the interests of reproducibility and transparency, all the code needed to
reproduce this work has been made freely available online at
\url{https://bitbucket.org/constantAmateur/binaryaccretion}.  See the readme
file in this repository for further details.

All figures in this paper were generated using the python package
{\sc matplotilb} \citep{MPL}.

\section{Acknowledgements}
\label{sec:ack}

We would like to thank Davide Gerosa, Giuseppe Lodato and Giovanni Rosotti for
useful discussions.  We thank Giovanni Rosotti for a critical reading and
comments on the manuscript.  We
acknowledge an anonymous referee for their comments which improved the manuscript.

We are indebted to Eduardo Delgado-Donate who was working on this problem
at the time of his untimely death in 2007.

This work has been supported by the DISCSIM project, grant agreement 341137 funded by 
the European Research Council under ERC-2013-ADG. Supercomputer time was
provided through DiRAC project grants DP022 \& DP047.

This work used the DIRAC Shared Memory Processing system at the University of
Cambridge, operated by the COSMOS Project at the Department of Applied
Mathematics and Theoretical Physics on behalf of the STFC DiRAC HPC Facility
(\url{www.dirac.ac.uk}). This equipment was funded by BIS National E-infrastructure
capital grant ST/J005673/1, STFC capital grant ST/H008586/1, and STFC DiRAC
Operations grant ST/K00333X/1. DiRAC is part of the National
E-Infrastructure.

This work used the DiRAC Data Analytic system at the University of Cambridge,
operated by the University of Cambridge High Performance Computing Service on
behalf of the STFC DiRAC HPC Facility (\url{www.dirac.ac.uk}). This equipment was
funded by BIS National E-infrastructure capital grant (ST/K001590/1), STFC
capital grants ST/H008861/1 and ST/H00887X/1, and STFC DiRAC Operations grant
ST/K00333X/1. DiRAC is part of the National E-Infrastructure.

\bibliographystyle{mn2e}
\bibliography{references}

\end{document}